# Polymer nanoreactors shield perovskite nanocrystals from degradation

Verena A. Hintermayr[1,2,‡], Carola Lampe[2,3,‡], Maximilian Löw[1,2,‡,†], Janina Roemer[2,4], Willem Vanderlinden[5], Moritz Gramlich[2,3], Anton X. Böhm[2,4], Cornelia Sattler[2,4], Bert Nickel[2,4,*], Theobald Lohmüller[1,2,*], and Alexander S. Urban[2,3,*]

1 Chair for Photonics and Optoelectronics, Nano-Institute Munich, Department of Physics, Ludwig-Maximilians-Universität München, Königinstr. 10, 80539 Munich, Germany

2 Nanosystems Initiative Munich (NIM) and Center for NanoScience (CeNS), Schellingstr. 4, 80799 Munich, Germany

3 Nanospectroscopy Group, Nano-Institute Munich, Department of Physics, Ludwig-Maximilians-Universität München, Königinstr. 10, 80539 Munich, Germany

4 Soft Condensed Matter Group, Department of Physics and Center for Nanoscience (CeNS), Ludwig-Maximilians-Universität, Geschwister-Scholl-Platz 1, 80539 Munich, Germany

5 Chair for Applied Physics, Department of Physics and Center for NanoScience (CeNS), Ludwig-Maximilians-Universität München, Amalienstr. 54, 80799 Munich, Germany

ABSTRACT

Halide perovskite nanocrystals (NCs) have shown impressive advances, exhibiting optical properties that outpace conventional semiconductor NCs, such as near-unity quantum yields and ultrafast radiative decay rates. Nevertheless, the NCs suffer even more from stability problems at ambient conditions and due to moisture than their bulk counterparts. Herein, we report a strategy of employing polymer micelles as nanoreactors for the synthesis of methylammonium lead trihalide perovskite NCs. Encapsulated by this polymer shell, the NCs display strong stability against water degradation and halide ion migration. Thin films comprising these NCs exhibit a more than 15 fold increase in lifespan in comparison to unprotected NCs in ambient conditions and even survive over 75 days of complete immersion in water. Furthermore, the NCs, which exhibit quantum yields of up to 55% and tunability of the emission wavelength throughout the visible range, show no signs of halide ion exchange. Additionally, heterostructures of MAPI and MAPBr NC layers exhibit efficient Förster resonance energy transfer (FRET), revealing a strategy for optoelectronic integration.

INTRODUCTION

Halide perovskite nanocrystals (NCs) were first realized in 2014[1] and since then have been synthesized through many different procedures and studied in detail with a focus on morphology, optical and electrical properties.[2-7] They have been highly optimized to enable bright, tunable photoluminescence (PL) emission throughout the entire visible range for use in lighting applications.[8, 9] Furthermore, their size and shape can be varied from bulk-like 3D NCs, to 2D nanoplatelets (NPls), 1D nanowires (NWs) and nanorods (NRs) and even quasi-0D quantum dots (QDs).[10-16] Despite these achievements and a plethora of studies, perovskite NCs still exhibit severe limitations for an unrestricted use in optoelectronic applications. Akin to their bulk counterparts, the NCs degrade due to external environmental influences such as humidity, heat and ultraviolet (UV) light illumination.[8, 17, 18] Strategies to mitigate water-induced degradation of perovskites have often focused on encapsulating entire working devices in water-impermeable materials[19] or underneath 2D-perovskite layers, which are less prone to moisture-induced degradation.[20] Particularly hydrophobic organic ligands and polymers were suggested to enhance moisture resistance.[21, 22] Alternatively, NCs have been synthesized inside solid matrices, such as $SiO_2$, alumina, or high-molecular weight polymers.[23-25] This approach, however, leaves the NCs fixed inside the matrix such that they cannot be assembled subsequently in defined structures such as highly uniform emitting layers, nor can they be investigated individually. Another unique property of halide perovskites is that the halide ions are extremely mobile inside the perovskite crystal structure, facilitating a rapid exchange of the entire halide ion content.[26-29] This has been used to tune the PL emission of perovskite NCs with specific geometries that can only be synthesized directly with a specific halide ion, typically bromide.[30, 31] However, this effect is not only beneficial. In light-emitting diodes (LEDs) comprising mixed halide content, large applied voltages cause the halide ions to migrate, inducing halide-phase segregation.[32] This results in unwanted shifts of the PL emission during device operation. Stability against halide ion migration has only been scarcely explored so far. Manna and coworkers subjected a NC film to X-ray radiation, causing intramolecular bonding between the organic ligands coating the NCs, leaving them impervious to halide ion migration and slightly enhancing their stability against water-induced degradation.[18] Obviously, this approach is not feasible for upscaling to mass production of devices. Thus, a method is required that i) enables a synthesis of perovskite NCs of controllable size and emission wavelength ii) allows for easy incorporation of the NCs into electronic devices and iii) prevents water-induced degradation and halide ion migration.

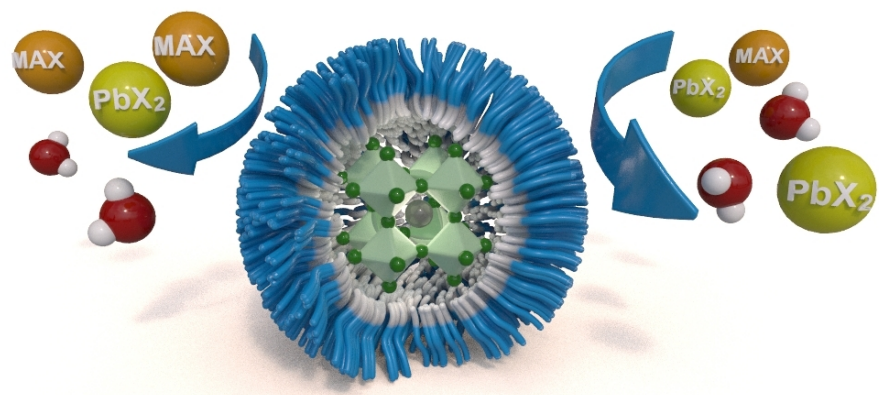

**Figure 1.** Scheme of perovskite nanocrystal encapsulation via diblock copolymer micelles to mitigate moisture-induced degradation and halide ion migration.

In this work, we report on such a strategy (**Figure 1**). We show a direct single step synthesis of perovskite NCs by means of diblock-copolymer micelles, which serve as nanoreactors for the formation of perovskite NCs and encapsulate them, vastly improving their stability. Block copolymers have been used for the synthesis of uniform metallic or metal oxide nanoparticles. The advantage of this method lies in the fact that the particles display a high monodispersity, which is unmatched compared to any other method. Furthermore, the size of the created nanoparticles can be adjusted with nearly atomic precision, which renders it possible to study and compare, for example, catalytic properties of sub-10nm particles with great accuracy.[33] In the case of noble metal particles, nanocrystal formation requires an additional processing step, either by chemical reduction or by plasma treatment.[34-36] The later serves the additional purpose of removing the polymer shell that is surrounding the final particles. In the case of metal oxide particles or semiconductor NC formation, such a chemical reduction step might not be required or even be necessary. Additionally, these highly ordered thin films can be easily incorporated into complex heterostructures of multiple NC species, either through deposition into predefined patterns or through post-processing, for example by e-beam lithography.[37] Recently, an attempt was made to also apply this technique to all-inorganic perovskites.[38] However, water-induced degradation was only slightly reduced for a matter of hours and halide ion migration was not demonstrated. We show that during our synthesis the precursor salts diffuse into the cores of the micelles, where halide perovskite NCs spontaneously crystallize. These exhibit strong photoluminescence as documented by quantum yields of up to 55%. Notably, this procedure does not require a secondary step to induce the crystallization. More importantly, the micellar-embedded NCs are vastly superior to standard halide perovskites in terms of stability against humidity. Not only were NC films strongly emissive after more than 200 days being exposed to ambient conditions, but they also exhibited fluorescence after 75 days of complete submersion in water. Additionally, no halide ion migration occurred in such films. The dynamics of the block-copolymer system also enable a fine-tuning of NC size, spacing and even shape. Energy transfer from bromide- to iodide-containing micelle-encapsulated NCs via Förster resonance energy transfer (FRET) reveals an approach for optoelectronic integration. This work constitutes a new approach to synthesizing perovskite NCs of controllable size and composition, with vastly improved resistance to halide ion migration and environmentally-induced degradation. Not only can this be expected to advance long-term durability and stability of optoelectronic applications, but the approach is also promising for realizing novel structures and technological principles, such as perovskite NC energy funnels,[39] which otherwise would not be possible.

RESULTS

Perovskite NCs were obtained through a diblock copolymer-templated wet synthesis, adapted from literature and detailed in the methods section.[37] In short, a polystyrene-poly(2-vinlypyridine) (PS-b-P2VP) diblock copolymer was added to toluene, a frequently used antisolvent for halide perovskites, where the polymer spontaneously forms core/shell micelles with the P2VP part forming the core and the PS part forming the shell. Perovskite precursor salts ($PbX_2$ and MAX with X= Cl, Br, I and MA= methylammonium) added to the dispersion, accumulate in the cores of the micelles due to diffusion. Here, the solubility product is changed and the precursors crystallize to form perovskite, an entropically driven process. This way, the core confines the volume available for perovskite formation, while the shell separates individual reservoirs.

The following data was obtained for a polymer with 266 units of PS and 41 units of PVP ($PS_{266}$-$PVP_{41}$), which has been shown to form stable core/shell micelles in toluene.[40, 41] Additionally, aromatic nitrogen-containing molecules have been shown to control the crystallization speed of perovskites[42] and have a passivating effect on the perovskite crystals. We thus expect the pyridine groups to act as crystallization centers for the perovskite formation inside the micelles and enhance their optical properties.[43] Consequently, the micelles act as nanoreactors, enabling the precursor ions to enter the cores, where they then crystallize to form perovskite NCs. The optical properties of the NC dispersions were investigated by PL and absorption spectroscopy (cf. Figure S1). Importantly, as shown in Figure 2a, the dispersions exhibit a single photoluminescence peak, whose maximum can be tuned throughout the visible range (from 400 nm to 760 nm) by varying the halide composition in the precursor salts. These spectral positions match those obtained in previous publications for bulk-like perovskite NCs, indicating no or only very weak confinement and consequently NC sizes larger than the bulk excitonic Bohr radii of the perovskites of approximately 3-5 nm.[45] The emission spectra are very narrow with full width at half maximum (FWHM) values between 80 meV and 100 meV for all $Br_xCl_{3-x}$ mixtures and the pure MAPI, comparing favorably to all-inorganic perovskite nanocrystals.[6] Only the $Br_xI_{3-x}$ mixtures exhibit wider, slightly asymmetric spectra with FWHM values ranging from 125 meV to 215 meV and shoulders extending to larger energies. This could be a result of a non-uniform halide content within the NC ensembles or of quantum-confinement effects induced by the polydispersity of the NCs. The NCs are very efficient as indicated by quantum yield (QY) values of up to 63 % for the pure bromide and 55% for the pure iodide samples and slightly lower values for the mixed halide samples. Importantly, the dispersions are stable over time, with PL and absorption spectra not changing discernibly over several months.

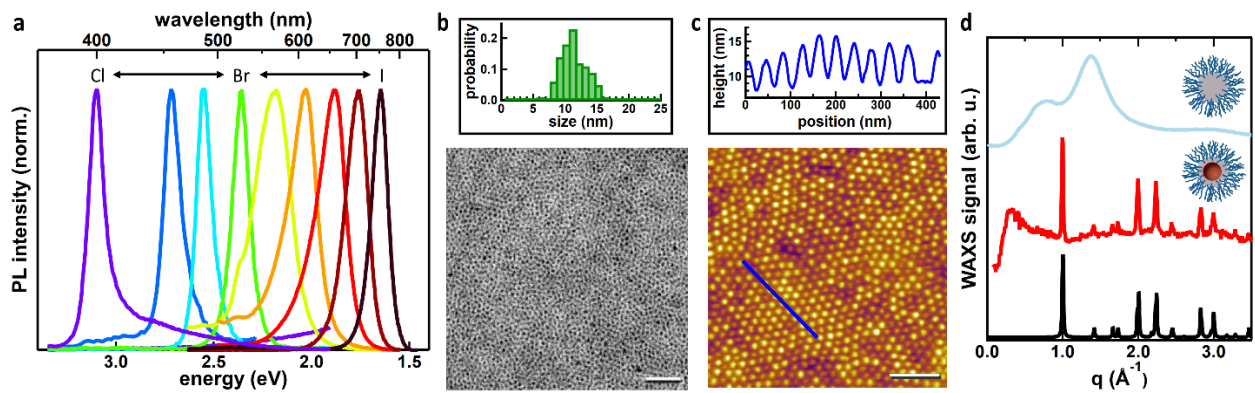

**Figure 2. Optical and morphological characterization of nanoreactors-encapsulated perovskite NCs. a)** Photoluminescence (PL) spectra of $MAPbX_3$-NCs with the halide composition varying from Cl to Br to I. The emission maximum blueshifts concomitantly from 400 to 760 nm. **b)** Scanning electron microscopy image of a monolayer of NCs deposited on a substrate. The NCs exhibit monodisperse size distribution and spacing and consequently form highly dense films. **c)** Atomic force microscopy (AFM) confirms the high quality of films with a surface coverage of 99.5% over large areas ($cm^2$). A scan over several micelles shows highly regular spacing and a dip between the micelles of 4-5 nm. **d)** Wide-angle X-ray scattering (WAXS) provides insights into the micelle formation and subsequent loading with perovskite NCs. For the MAPI NCs, the WAXS signal strongly resembles that obtained for bulk MAPI crystals, as per Stoumpos et al.[44]

In order to confirm the formation of micelles and that they are preserved upon addition of the precursor salts, we performed small angle neutron scattering (SANS) experiments using deuterated toluene (d-toluene). As shown in Figure S2 of the Supporting Information, the drop of intensity at higher q, described well by a power law $I(q) \sim q^{-1.7}$, signifies scattering from swollen chains in an ideal solvent, confirming that toluene acts as a selective solvent for the PS shell. This is retained in the sample after precursor salt addition, while a pronounced increase of scattering at smaller q values, with a pronounced double peak signifies a core/shell particle, likely with the perovskite accumulated in the cores. A model free analysis of the micellar size is possible by Gunier law yielding a radius of gyration of $R_G$=16 nm. To investigate the cores in more detail, we employ transmission electron microscopy (TEM) on thin films of the micelles. Images show a very strong scattering from the cores, revealing that the NCs are highly homogeneous with sizes of 11 ± 2 nm and center-to-center distances of 27 ± 4 nm (Figure 2b). By varying the properties of the block copolymer used, we could tune both the size (from 27 ± 2 nm to 6 ± 1 nm) and the spacing of the NCs (40 ± 3 nm to 11 ± 2 nm ) in a wide range (see Supporting Information Figures S3, S4). Due to the high homogeneity, it was possible to fabricate high quality thin films over very large areas up to cm$^2$ by use of the dip-coating method. The films were probed with atomic force microscopy (AFM), as shown for the case of a monolayer film in Figure 2c. We measured a surface coverage of 99.5% with a typical surface roughness of less than 2 nm (see Figure S5). By measuring at the edge of the deposited polymer-encapsulated NC film on the substrate, we determine the height of it to be 12 ± 2 nm matching the value determined from the TEM measurements (cf. Figure S6). This suggests that the polymer is compacted and forms only a very thin layer above and below the NCs.

Small angle X-ray scattering (SAXS) on drop-casted samples of empty micelles reveals a pronounced peak (q=0.0315 Ang-1) characteristic for close sphere packing with a sphere diameter of d=22.4 nm (cf. Figure S7, solid disks.) The SAXS intensity is largely increased upon halide intercalation (cf. Figure S7, open squares), while the peak position remains rather unchanged. This suggests that the diameter of the dry micelles remains similar, or even slightly condenses, after loading of the core with precursor salts and matches closely with the TEM and AFM results. We employed wide-angle X-ray scattering (WAXS) experiments to corroborate our findings and confirm the formation of perovskite NCs. For this, we dropcasted the empty micelles and those containing perovskite NCs onto parylene foil and performed the experiments in transmission geometry using a Mo X-ray lab source (see methods section for the details). The diffraction profile of the empty micelles is shown in Fig. 2d as the light blue curve. The broad diffraction at q= 0.8 Å-1 and 1.4 Å-1 closely resembles the diffraction of bulk PS and constitutes the fingerprint of the empty micelles.[46] Upon incorporation of the perovskite precursors into the micelles, the diffraction pattern changed dramatically, shown as the red curve. Here, the diffraction profile for crystalline MAPI is recovered, as shown in data previously obtained for bulk MAPI crystals by

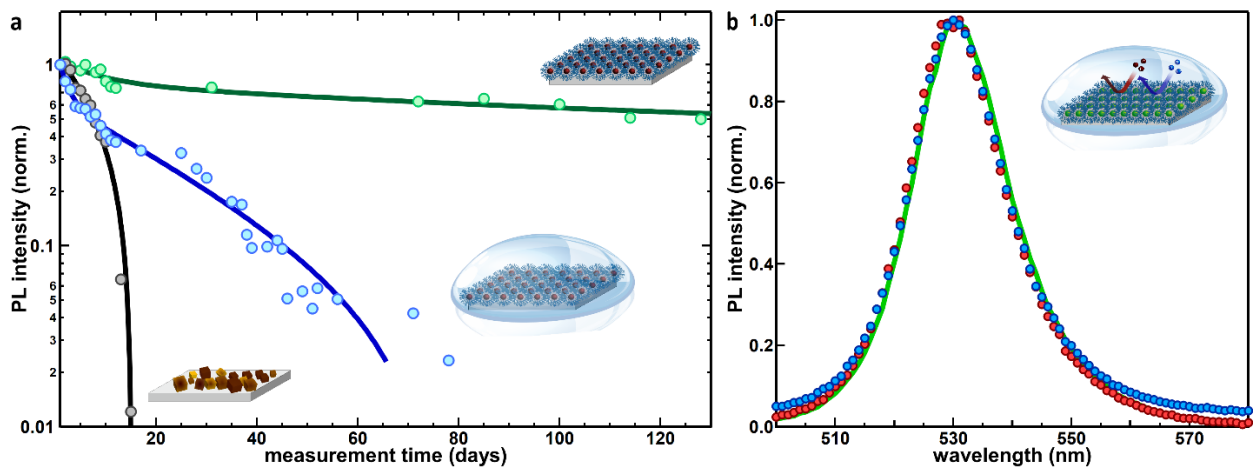

**Figure 3. Enhanced stability of diblock copolymer-encapsulated perovskite NCs. a)** Temporal development of PL intensity of perovskite NC films. Reference MAPI NCs synthesized according to Hintermayr et al. (black curve) degrade in ambient conditions completely within 13 days. In contrast, the encapsulated NCs (green curve) retain nearly 60% of the initial PL intensity after 130 days. Even completely submersed in water (blue curve), the encapsulated NCs exhibit discernible PL for over 75 days. b) PL spectra of a film comprising encapsulated MAPBr NCs (green line) and of the same films subjected to aqueous solutions of lead halide (blue points: $PbCl_2$, red points: $PbI_2$). As there is nearly no difference between the spectra, the polymer nanoreactors clearly prevent halide ion migration into or out of the micelles.

Stoumpos et al (black curve).[44] This confirms that the material within the polymer micelles is in fact bulk-like MAPI perovskite.

The main goal of this work was to synthesize high quality, controllable perovskite NCs that were protected from environmentally-induced degradation (water, oxygen, heat, UV-light) and halide ion migration. To determine the effectiveness of our system we fabricated several identical thin films comprising $PS_{266}$-$PVP_{41}$-coated MAPI-NCs using dropcasting to obtain films thick enough to quantify PL emission reliably. Additionally, reference samples comprising bulk MAPI nanocrystals, were synthesized as previously described and compared to the polymer encapsulated-perovskite NCs.[11] A first set of samples was stored in ambient conditions, *i.e.* under normal daylight illumination and with a relative humidity of approximately 40 %. Reference unprotected NCs show a rapid blueshift of the PL emission maximum and a simultaneous decrease of the PL intensity to zero on day 13 (black curve, Figure 3a; cf. Figure S8). This indicates that the unprotected NCs degrade rapidly from the outside in, leading to smaller NCs exhibiting quantum confinement and reduced PL emission and ultimately complete degradation. In contrast, the shape and position of the PL spectra for the polymer-encapsulated NCs do not change and the PL intensity decreases substantially slower (blue curve, Figure 3a). After 150 days, the PL intensity retained over 50% of the original value. The PL intensity decrease becomes even slower with time and the PL intensity seemingly leveled out at ~ 40%, even after over 220 days of measurements (see Figure S9). Consequently, the NC lifetime in ambient conditions is increased by a factor of more than 15 times in comparison to unprotected NCs. Going a step further, we immersed the polymer-encapsulated NC films into water, tracking their PL emission. While reference samples degrade instantaneously, the degradation is much slower in the micellar-protected sample. Displaying a slightly faster decay than the sample in only ambient conditions, the immersed NCs retained more than 40% of the original PL intensity after 13 days and even exhibited a discernible PL signal after more than 75 days of complete immersion. Clearly, the polymer shell significantly enhances the stability of the perovskite NCs to degradation from water exposure, even under full immersion in water.

To investigate how the polymer micelles affect the migration of halide ions, we synthesized three additional encapsulated NC dispersions, one with each type of halide ion (Cl, Br, I). These were deposited on substrates through dropcasting and submerged into aqueous solutions containing the respective other two halides (e.g. $PbCl_2$ and $PbI_2$ for $MAPbBr_3$-NCs). PL spectra of the films showed that there was no noteworthy change even after several days, as demonstrated here for the case of the MAPbBr$_3$-NCs (Figure 3b). This suggests that the polymer micelle is impermeable to halide ions in a polar environment. These findings have important implications for device fabrication, as they suggest that once the NCs are integrated into devices, the polymer shielding prevents the migration of ions and thus should enable a stable, spectrally constant emission.

While the polymer shielding has proven to be effective for stabilizing the NCs, obviously this might be a deterrent for optoelectronic integration. The easiest form of integration is as color filters in a standard liquid crystal display (LCD) scheme with the NCs - as downconverters - absorbing light of a UV or blue LED and reemitting blue/green/red light.[47, 48] Here, the color purity and extremely large modulation bandwidth make perovskite NCs an excellent choice. Ideally one would like to transition to a full LED which requires charge carrier injection into the emissive NCs. Direct charge transfer typically can only deal with non-conductive spacings of less than one nanometer, and so in this configuration is unlikely to occur. The polymer surrounding the nanocrystals would likely be needed to rendered conductive in order to enable this. However, this is not the only strategy for injecting charge carriers into NCs. Förster radiative energy transfer or FRET for short, is a mechanism by which a donor entity can exchange energy, provided the emission of the donor and the absorption of the acceptor overlap and the transition dipole moments overlap. In this method, distances of up to 10 nm have been shown to still enable FRET.[49, 50] In the systems we have produced here, we have shown that the gaps between the NCs can easily be held below 10nm and so should enable FRET. To investigate this, we synthesized MAPI and MAPBr NCs inside the PS$_{266}$-PVP$_{41}$ polymer micelles. A sample was prepared by dropcasting a thick micelle-encapsulated MAPBr NC film onto a silicon substrate and then spincoating a very thin film of micelle-encapsulated MAPI NCs on top (Figure 4a). Excited with a laser at 450 nm, both types of NCs exhibit PL while the MAPBr NCs can potentially transfer their energy via FRET to the MAPI NCs (Figure 4b). As shown in Figure 4c, the PL spectra from pure films of each NC type (MAPBr: green, MAPI: red) correspond nearly perfectly to the PL spectrum emitted from the combined sample (yellow). Importantly, the PL peak of the MAPI component is nearly as strong as that of the MAPBr component, despite being significantly thinner. However, in order to verify energy transfer, we compare the PL decay of the MAPBr component both in the pure film and in the combined sample (Figure 4d). In both cases, we observe a decay, which is of multiexponential origin, however, it is significantly faster in the combined film. Taking the time at which the PL intensity has fallen to 1/e, the PL lifetime decreases from 0.59 ns to 0.41 ns. This means that an additional decay pathway for the MAPBr is present in the mixed sample, which we attribute to FRET-mediated energy transfer. The transfer efficiency can be obtained through the lifetimes of the sample and is given by: $\eta_{FRET} = 1 - \frac{\tau'_D}{\tau_D}$ with the PL lifetimes of the pure sample, $\tau_D$ and of the mixed sample $\tau'_D$.[51] In this case, we obtain a transfer efficiency of 30.5%, which is remarkable considering the extremely thin layer of MAPI NCs. This confirms that energy transfer can occur between micelle-encapsulated NCs, allowing for optoelectronic integration and enabling novel nanostructures such as cascaded energy transfer systems or energy funnels.

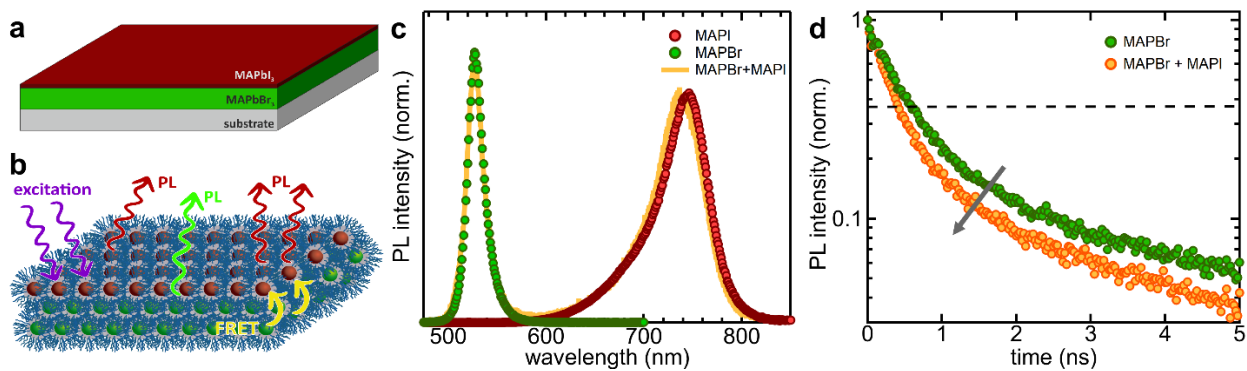

**Figure 4. Nonradiative energy transfer between NCs of different composition. a)** Scheme of the experimental structure with a thin spin-coated layer of MAPI NCs on top of a thick layer of MAPBr NCs. **b)** Scheme depicting excitation, emission and energy transfer in the sample. **c)** PL spectra of the pure MAPBr sample (green), the pure MAPI sample (red) and the combined structure (yellow). **d)** PL decay of the pure MAPBr sample (green) and of the MAPBr in the combined structure (yellow) showing an increased decay rate.

**CONCLUSIONS**

In summary, we have synthesized perovskite NCs using diblock copolymer nanoreactors as a growth template. Importantly, the precursor salts accumulate in the cores of the micelles where they instantaneously begin to crystallize to form the perovskite without the need for a secondary reduction step. The process is highly reproducible, resulting in monodisperse luminescent nanocrystals of arbitrary halide composition with sizes tunable through the length of the core polymer block. The nanoreactor-NCs are very stable, far outlasting unprotected NCs in ambient conditions, even retaining substantial PL emission even after 75 days of complete immersion in water. Moreover, the polymer shield completely prevents halide ion exchange in aqueous solvents. This and the NCs' ability to self-assemble in near-perfect films on substrates promises to enable the integration into optoelectronic devices. FRET between NCs of varying bandgap with an extremely high efficiency highlights a possible avenue for realizing this. Moreover, the block-copolymers should enable subsequent nanopatterning of thin films through photolithography and e-beam illumination, greatly expanding the range of possible device structures. Future studies will need to focus on the exact formation mechanisms and the specific roles of the polymer components as well as how to enable and optimize charge and/or energy transfer. Nevertheless, the proposed synthesis highlights an excellent strategy for mitigating the detrimental effects of environmentally-induced degradation and halide ion migration currently impeding the advancement of perovskite-based optoelectronics.

## ASSOCIATED CONTENT

**Supporting Information**. Materials and methods for the synthesis and characterization of the encapsulated perovskite NCs. Images of the micellar dispersions, PL spectra, TEM images and analysis of NCs, AFM analysis of NC films. The following files are available free of charge.

## AUTHOR INFORMATION

**Corresponding Author**

*Bert Nickel, Email: nickel@lmu.de

*Theobald Lohmüller, Email: t.lohmueller@lmu.de

*Alexander S. Urban, Email: urban@lmu.de

Present Addresses

†M.L: Max-Planck-Institut für Quantenoptik, Hans-Kopfermann-Str. 1, 85748 Garching, Germany

**Author Contributions**

‡ V.A.H, C.L. and M.L. contributed equally.

## ACKNOWLEDGMENT

This work was supported by the Bavarian State Ministry of Science, Research, and Arts through the grant "Solar Technologies go Hybrid (SolTech)", by the DFG through the grant "e-conversion" within the framework of the German Excellence Initiative and through SFB 1032 projects A07 and A08, by the European Research Council Horizon 2020 Marie Skłodowska-Curie Grant Agreement COMPASS (691185) and the ERC Grant Agreement PINNACLE (759744), and by the Research Foundation Flanders (W.V.). The authors would like to thank Markus Döblinger for helpful discussions concerning electron microscopy and Christoph Homann for realizing Figure 1. V.A.H, C.L., and M.L contributed equally to this work. T.L. and A.S.U contributed equally to this work

FOR TABLE OF CONTENTS ONLY

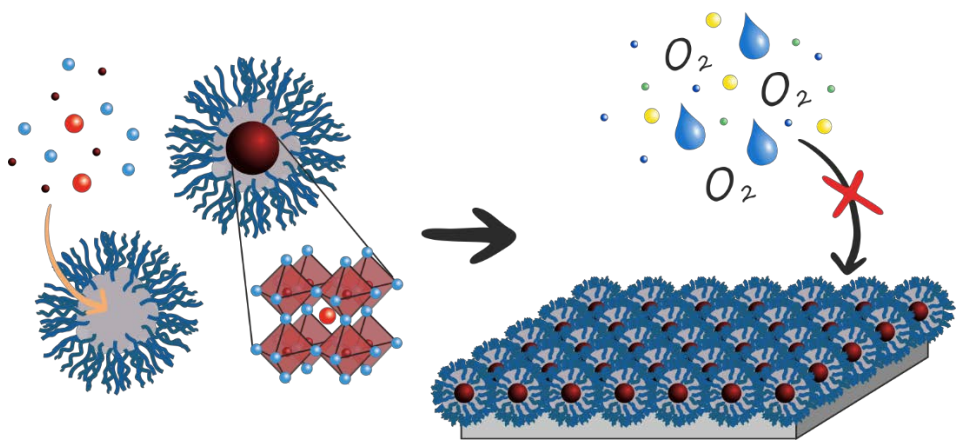

Supporting Information for

# Polymer nanoreactors shield perovskite nanocrystals from degradation

*Verena A. Hintermayr[1,2,‡], Carola Lampe[2,3,‡], Maximilian Löw[1,2,‡,†], Janina Roemer[2,4], Willem Vanderlinden[5], Moritz Gramlich[2,3], Anton X. Böhm[2,4], Cornelia Sattler[2,4], Bert Nickel[2,4,\*], Theobald Lohmüller[1,2,\*], and Alexander S. Urban[2,3,\*]*

1 Chair for Photonics and Optoelectronics, Nano-Institute Munich, Department of Physics, Ludwig-Maximilians-Universität München, Königinstr. 10, 80539 Munich, Germany

2 Nanosystems Initiative Munich (NIM) and Center for NanoScience (CeNS), Schellingstr. 4, 80799 Munich, Germany

3 Nanospectroscopy Group, Nano-Institute Munich, Department of Physics, Ludwig-Maximilians-Universität München, Königinstr. 10, 80539 Munich, Germany

4 Soft Condensed Matter Group, Department of Physics and Center for Nanoscience (CeNS), Ludwig-Maximilians-Universität, Geschwister-Scholl-Platz 1, 80539 Munich, Germany

5 Chair for Applied Physics, Department of Physics and Center for NanoScience (CeNS), Ludwig-Maximilians-Universität München, Amalienstr. 54, 80799 Munich, Germany

**Corresponding Author**

*Bert Nickel, Email: nickel@lmu.de
*Theobald Lohmüller, Email: t.lohmueller@lmu.de
*Alexander S. Urban, Email: urban@lmu.de

**METHODS**

**Formation of block copolymer-encapsulated perovskite NCs**

Perovskite NCs were obtained in a block copolymer-templated wet synthesis. The polymer ($1.56 \cdot 10^{-6}$ $mol$) was dissolved in toluene (5 ml) and stirred overnight to ensure complete formation of the micelles. Lead- and methylammonium halides were used as precursors. In order to grow crystals in the micelles, methylammonium halide (MAX, X = Cl, Br, I; $6.40 \cdot 10^{-5}$ $mol$) was added to the polymer solution and stirred vigorously for 5 hours. Afterwards, lead halide ($PbX_2$, X = Cl, Br, I; $6.40 \cdot 10^{-5}$ $mol$) was added and the solution was stirred again for at least 24 hours. In case of mixed halides $PbX_2$, MAX and MAY were used for $MAPbX_{3-x}Y_x$ (X, Y = Cl, Br, I) perovskites with the ratio 1 : 1 : 1-x. For further purification the suspension was centrifuged for 10 min at 5000 rpm. The supernatant was subsequently centrifuged a second time for 20 min at 12000 rpm. The resulting supernatant was analyzed further (cf. Figure S7).

**AFM imaging**

Atomic force microscopy imaging was performed on a commercial Multimode AFM, equipped with a Nanoscope III controller and a type E scanner. Images were recorded in amplitude-modulation mode on dried samples, under ambient conditions, and using Silicon cantilevers (Olympus; AC160TS; resonance frequency ~ 300 kHz). Typical scans were recorded at 1-3 Hz line frequency, with optimized feedback parameters and at 512*512 pixels. For image processing and analysis, Scanning Probe Image Processor (SPIP; v6.4; Image Metrology) was employed. Image processing involved background correction using global fitting with a third-order polynomial, and line-by-line correction through the histogram alignment routine.

### X-ray scattering

All X-ray scattering measurements were performed using a microfocus X-ray tube with a molybdenum anode. Samples were measured in transmission and the direct beam was blocked by a beamstop behind the sample to reduce scattering on air. The scattered intensity was recorded by a PILATUS 100K (Dectris Ltd) CMOS hybrid pixel detector mounted on a movable stage, allowing to measure up to $q = 3.5$ Å-1. The powder diffraction data was reduced to 1D using Igor Pro (Wave-Metrics) with the Nika package.[45] For each sample, a background measurement of a corresponding bare foil was subtracted. The powder simulation was performed using the software Mercury of the Cambridge Crystallographic Data Centre.[46, 47]

### Neutron scattering

The small-angle neutron scattering measurements were performed at SANS-1 at the Heinz Maier-Leibnitz Zentrum (MLZ). A detailed description of the SANS-1 is given in literature [49]. A circular aperture diameter of 14 mm was used. Clear beam, empty cuvette, and a water calibration measurement were performed to correct for transmission, detector efficiency, and geometry effects. The SANS data reduction was performed with BerSANS Software developed at the Hahn-Meitner-Institut (HMI). In order to cover the q-range, three different instrument configurations were used, varying in wavelength (4.5 Å and 12 Å ), collimation length (4m to 20 m) and sample-to-detector distance (2m to 20m).

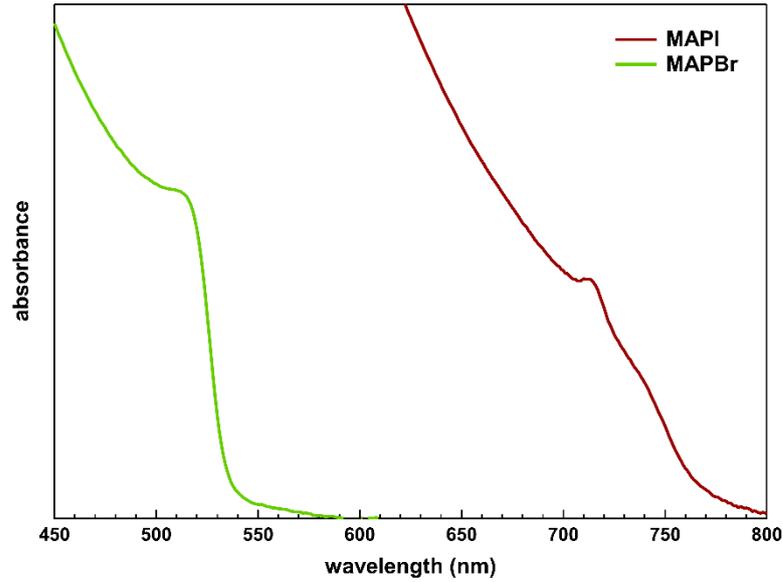

**Figure S1.** Absorption spectra of polymer micelle-encapsulated perovskite NCs comprising MAPI (red) and MAPBr (green).

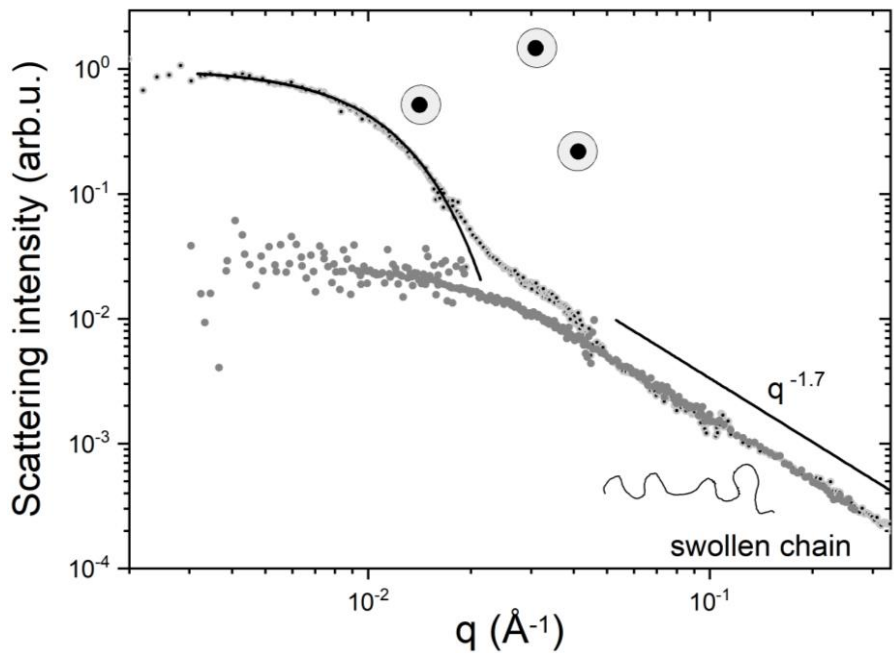

**Figure S2.** Small angle neutron scattering data. The SANS signal I(q) is shown as function of scattering vector q. The data are obtained from PS-b-P2VP dissolved in D-Toluol with and w/o addition of MAPI precursor salts, shown as core/shell disks and solid gray disks, respectively. For both samples, the intensity signal at $q>0.03$ Å$^{-1}$ drops according to a power law $I\sim q^{-1.7}$ as indicated by the black line, characteristically for scattering from swollen polymer chains. At $q <0.01$ Å$^{-1}$, an intensity plateau is observed, indicating well dispersed particles. The solid curve is a Guinier analysis of the initial drop at low q, i.e. $I(q)\sim\exp[-1/3\ (q*R_G)^2]$ ), in order to estimate the radius of gyration $R_G$ for the PS-b-P2VP micelles loaded with precursor salts. The curve is obtained for $R_G =16$ nm.

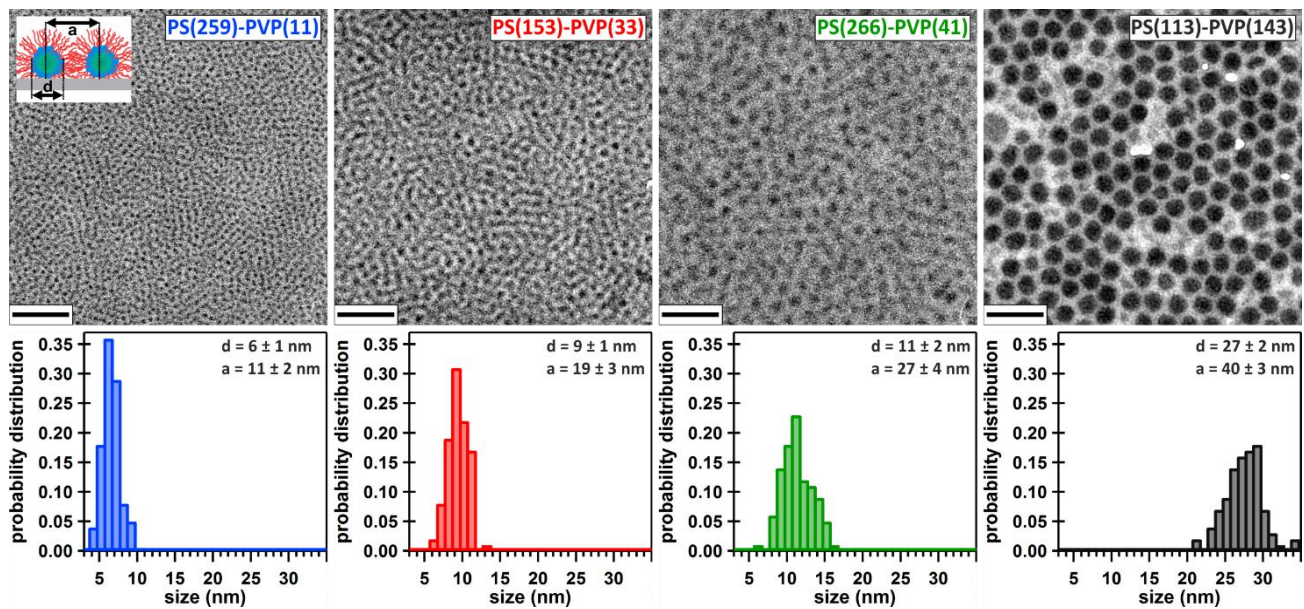

**Figure S3.** Size-tuning of encapsulated perovskite nanocrystals by varying the parameters of the block-copolymer. Both the size of the NCs and their spacing can be controlled by varying the lengths of the overall polymer and the polymer cores.

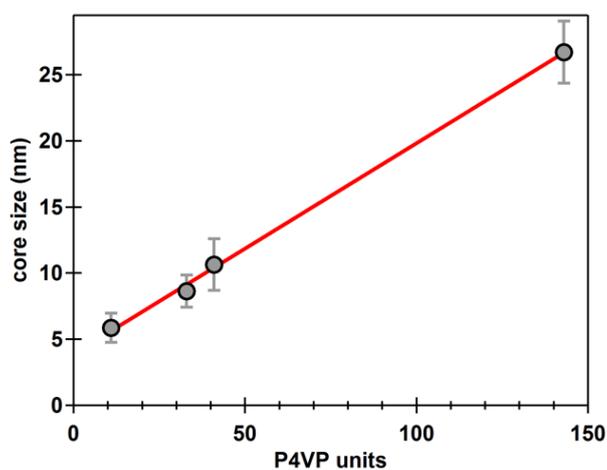

**Figure S4.** Dependence of the perovskite core size, as determined from TEM imaging, on the number of P2VP units within the block copolymer. A clear linear dependence is discernible.

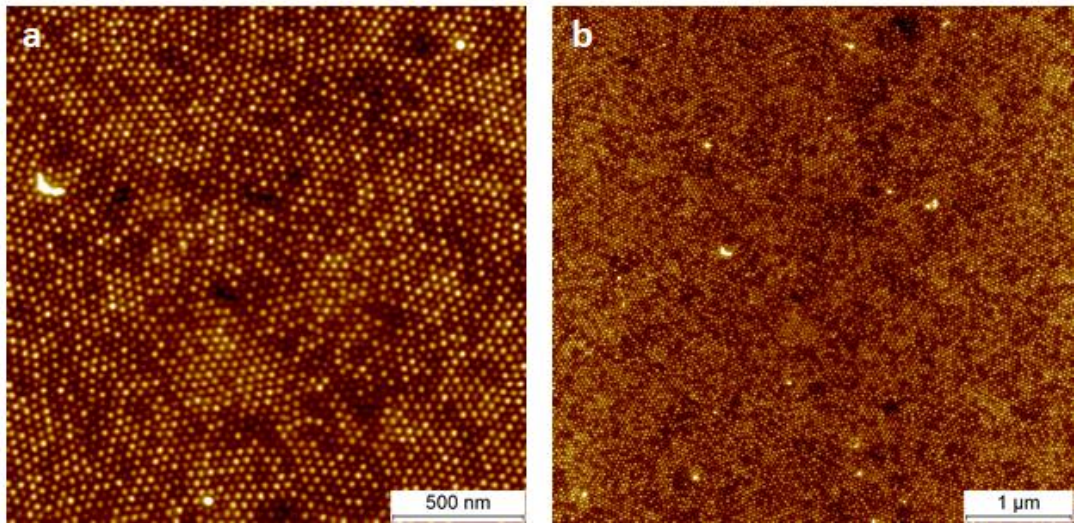

**Figure S5.** AFM images of polymer-encapsulated perovskite NCs films deposited through dipcoating showing a nearly perfect surface coverage with a monolayer of NCs.

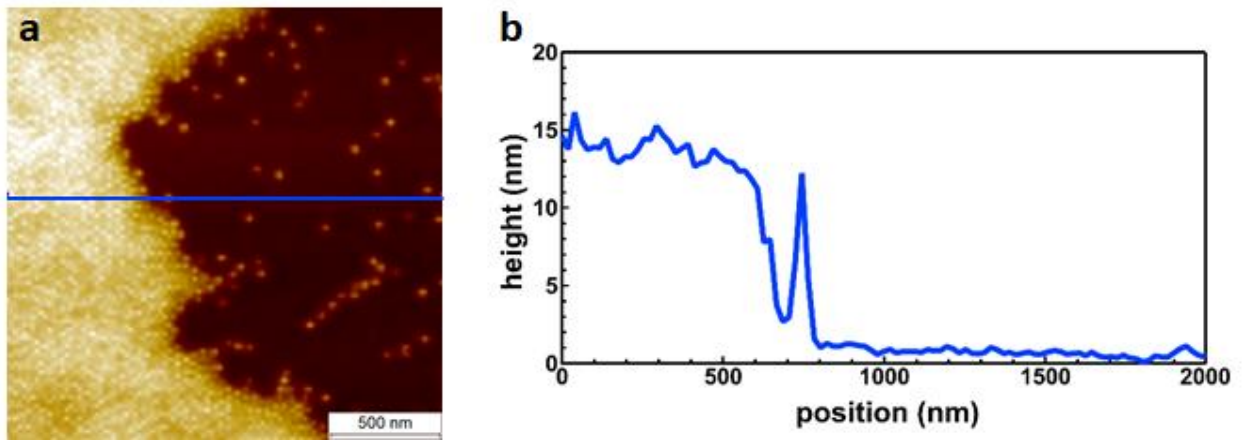

**Figure S6.** a) AFM image of the edge of the dipcoated polymer-encapsulated perovskite NC film. b) height profile of the film along the blue line in the AFM image. The scan shows a homogeneous film height of 12 ± 2 nm, corresponding to the TEM measurements of the NC sizes.

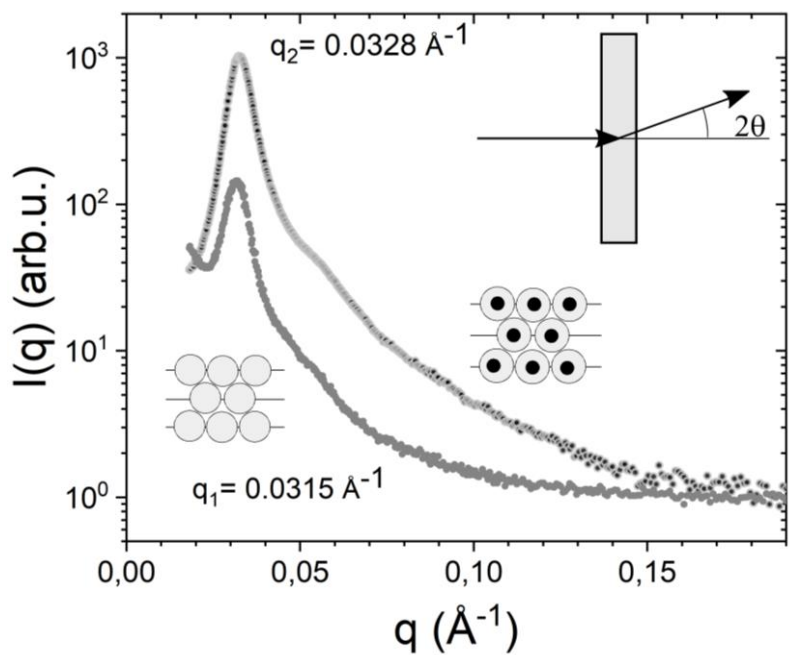

**Figure S7.** Small angle X-ray scattering data. The SAXS signal I(q) is shown as function of scattering vector q. The data are obtained from PS-b-P2VP dissolved in H-toluene with and w/o addition of MAPI precursor salts, after drop casting and drying, shown as core/shell and solid gray disks, respectively. The scattering geometry of the X-ray beam is shown in the inset. The intensities show a pronounced peak at $q_1$= 0.0315 Å$^{-1}$ and $q_2$=0.0328 Å$^{-1}$, respectively, for samples with and w/o precursor salt. The peak positions indicates a lattice spacing of $a_1$=19.1 nm and $a_2$=19.9 nm. This lattice spacing implies a sphere diameter of $d_1$=24.4 nm and $d_2$= 23.4 nm for close sphere packing, as shown schematically in the inset.

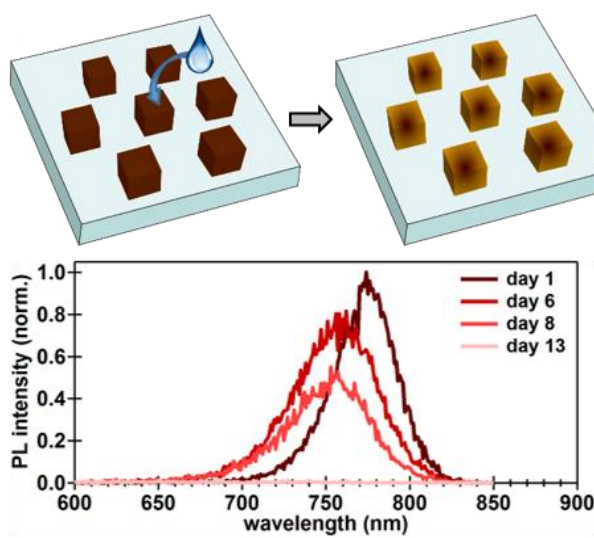

**Figure S8.** PL spectra of reference MAPI NCs deposited on substrates and left in ambient conditions for 13 days. The PL peak progressively blueshifts while the overall PL intensity is reduced until after 13 days no PL can be measured at all. This indicates a complete degradation of the NCs, likely due to susceptibility to moisture.

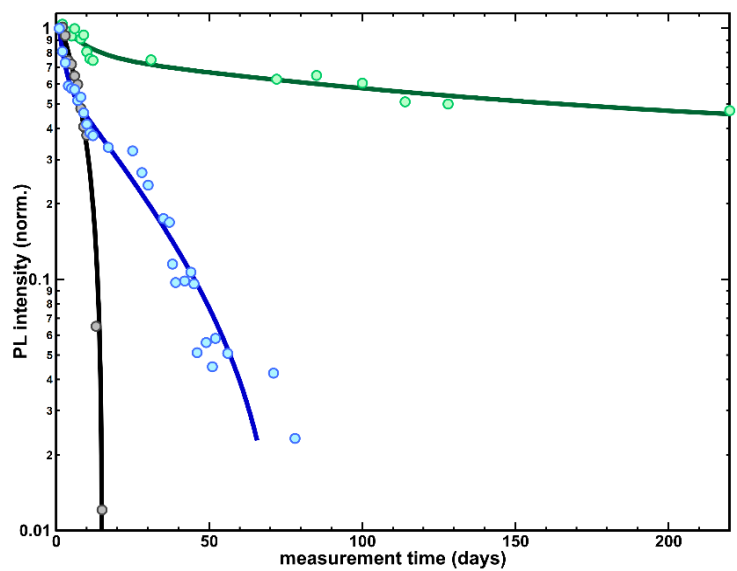

**Figure S9.** Long-term intensity measurements of perovskite NCs to monitor stability. While reference NCs degrade rapidly within 13 days (black curve) under ambient conditions (air environment, 40% humidty), polymer-encapsulated NCs retain more than 40% of their initial PL even after 220 days under the same conditions (green curve). Even after full-submersion in water, the polymer-encapsulated NCs show discernible PL emission after more than 75 days (blue curve).

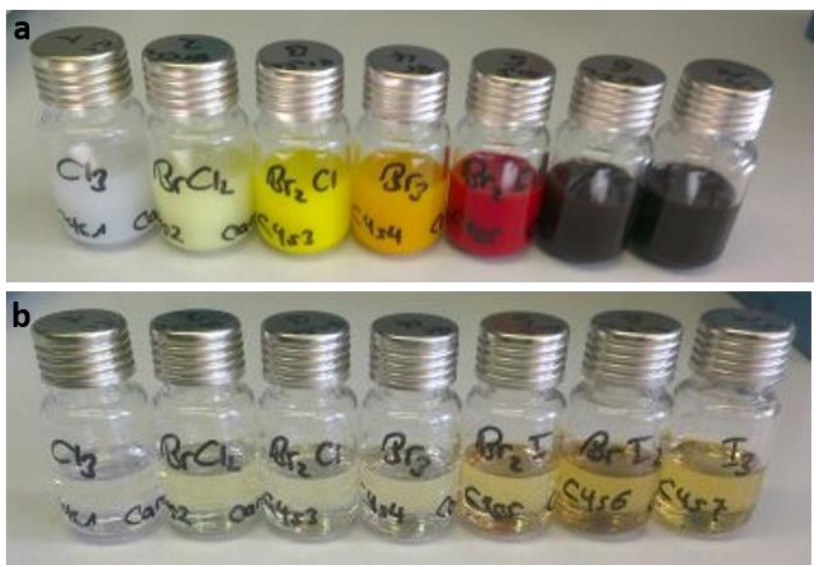

**Figure S10.** Block copolymer-encapsulated perovskite nanocrystal dispersions a) directly after synthesis and b) after centrifugation in order to remove larger crystals and residual polymer.